\pdfoutput=1
\documentclass[a4paper,twoside]{article}
\newcommand{\affil}[1]{$^{\rm #1}$}
\usepackage[authoryear]{natbib}
\bibpunct{(}{)}{;}{a}{}{,}
\usepackage{graphicx}
\usepackage[cmex10]{amsmath}
\usepackage{algorithmic}

\usepackage{array}

\usepackage{mdwmath}
\usepackage{mdwtab}

\usepackage[small]{caption}
\usepackage[tight,footnotesize]{subfigure}
\usepackage{fixltx2e}

\usepackage{upgreek}

\usepackage{url}

\date{} %Please leave the date blank
\title{\large\bf\flushleft Assessing The Accuracy Of Radio Astronomy Source Finding Algorithms}
\author{\parbox{\textwidth}{\flushleft
\vspace{-0.5cm}
{\it Stefan Westerlund\affil{A,B}, Christopher Harris\affil{A}, and Tobias Westmeier\affil{A}}\\
\vspace{0.4cm}
{\small \affil{A}\,ICRAR/University of Western Australia, M468 35 Stirling Highway, Crawley, WA 6009}\\
{\small \affil{B}\,Email: stefan.westerlund@icrar.org}}}
\begin{document}
\twocolumn[
\begin{changemargin}{.8cm}{.5cm}
\begin{minipage}{.9\textwidth}
\vspace{-1cm}
\maketitle
%
%
%%%%%%%%%%%%%     ABSTRACT    %%%%%%%%%%%%%
%Abstract of no more than 200 words here.
\small{\bf Abstract:}
This work presents a method for determining the accuracy of a source finder algorithm for spectral line radio astronomy data and the Source Finder Accuracy Evaluator (SFAE), a program that implements this method. The accuracy of a source finder is defined in terms of its completeness, reliability, and accuracy of the parameterisation of the sources that were found. These values are calculated by executing the source finder on an image with a known source catalogue, then comparing the output of the source finder to the known catalogue. The intended uses of SFAE include determining the most accurate source finders for use in a survey, determining the types of radio sources a particular source finder is capable of accurately locating, and identifying optimum  parameters and areas of improvement for these algorithms. This paper demonstrates a sample of accuracy information that can be obtained through this method, using a simulated ASKAP data cube and the \textsc{Duchamp} source finder.

%%%%%%%%%%%%%     KEYWORDS    %%%%%%%%%%%%%
\medskip{\bf Keywords:} methods: data analysis
% Please write all keywords in lower case. PASA uses the
% standard list of subject headings adopted by The Astrophysical Journal
% and available from http://www.journals.uchicago.edu/ApJ/keywords_text.html.
% Keywords are separated by em-dashes, i.e. ---

%%%%%%%%DO NOT EDIT%%%%%%%%%%%%
\medskip
\medskip
\end{minipage}
\end{changemargin}
]
\small
%Radio Astronomy; Source Finders; Completeness; Reliability;

\section{Introduction}
A critical aspect of astronomy is the quantitative analysis of objects such as stars and galaxies. Astronomers search through data collected from telescopes to determine the location and properties of these objects. A common technique is to use a computer program to search astronomical data, followed by manual inspection to confirm sources of electromagnetic radiation. However, using people to visually search telescope data is time-consuming and expensive.

Telescope installations that will be operational in the near future, such as the Australian Square Kilometre Array Pathfinder (ASKAP) or the Square Kilometre Array (SKA), will produce orders of magnitude more data than previous telescopes \citep{ASKAPDataOutput}. Partially or completely manual searching techniques will not scale to handle such a large volume of information. Therefore a purely automated approach is needed. As part of designing an algorithm that accurately and efficiently searches large amounts of spectral line telescope data, it is necessary to determine the accuracy of that algorithm. This paper describes the method used by the Source Finder Accuracy Evaluator, a program which measures the accuracy of a source finding program.

\section{Background}
A \emph{source finder} is a program that searches radio astronomy data, in the form of a data cube, and returns a catalogue of the sources it finds, with the parameters of each source, such as the objects' position and flux. Surveys such as HIPASS use a combination of automated source finding programs and manual inspection to provide a source list. In the case of HIPASS, two source finding programs, \url{MultiFind} and \url{TopHat} were run on the telescope data and their candidate catalogues were merged and examined manually to confirm the detected objects \citep{HIPASSCatalogue}.

Radio telescope spectral line data is fed into a source finder program as a data structure called a \emph{data cube}. A data cube can be thought of as a three-dimensional grid, where the three dimensions usually are right ascension, declination and either frequency or velocity. Each cell contains the flux for the area in the sky and the spectral range represented by that cell. New telescopes, such as ASKAP, will produce orders of magnitude more data than previous installations. For example, the HIPASS survey used the existing Parkes telescope to produce a 111 MB data cube every ten hours \citep{HIPASSDataReduction}, for a data output rate of 24.7 Kbps. By comparison, ASKAP is capable of producing a data cube of as much as 4 TB in size every eight hours, depending on the configuration \citep{ASKAPDataOutput}, for a data output rate of 1.14 Gbps, an increase of over four orders of magnitude compared to HIPASS.

ASKAP's increased data rate will require faster source finding programs in order to be able to process the data fast enough to keep up. In order to achieve the required data processing rate, source finders will need to exploit the computational power of modern supercomputers. Previous work \citep{DuchampParallel} shows that in order to do this, an efficient parallel implementation of the source finding algorithm must be made. Additionally, such data sets will contain many more sources than previous surveys, such that manual confirmation of all the sources will require an extremely large amount of work. Therefore, in order for the data to be processed in a reasonable amount of time and effort from observers, the accuracy of the automated source finder must be sufficiently high that manual confirmation is unnecessary.

A source finder's accuracy has three aspects: completeness, reliability and parameter correctness. Completeness and reliability describe the accuracy with which the source finder has found objects in the data set, as noted by \citet{HIPASSAccuracy}. \emph{Completeness} is the portion of objects in the data cube that are found by the source finder program. If $n_d$ is the number of sources present in the data cube and $n_r$ is the number of real sources that have been located by the source finder, then the completeness, $C$, can be calculated using the following expression:
\begin{equation}
	\label{CEq} C = \frac{n_r}{n_d}
\end{equation}
\emph{Reliability} is the portion of the detections from the source finder that exist in the data cube. The reliability, $R$, can be calculated in a similar manner to the completeness. If $n_t$ is the number of all the objects that the source finder has located, both true and false positives, then the reliability can be calculated using the expression:
\begin{equation}
	\label{REq} R = \frac{n_r}{n_t}
\end{equation}

It is often more useful to describe these values as a function of the parameters of the sources, in order to analyse how the source finder performs for different types of sources. In particular, the completeness and reliability of the sources are often calculated as a function of peak flux, integrated flux, signal to noise ratio (SNR), or spectral width. The completeness and reliability can be calculated by binning the sources by the parameter in question, or by calculating the cumulative completeness and accuracy. The binning function must have enough bins to show the distribution of sources, but not so many as to under-sample the bins. Using a suitable binning function, the completeness of bin $i$, $C_{i}$, and the reliability of bin $i$, $R_{i}$ can be calculated using the number of real sources that have been found by the source finder in bin $i$, $n_{r,i}$, the number of sources in the data set that are in the $i$th bin, $n_{d,i}$ and the total number of sources found by the source finder that are in bin $i$, $n_{t,i}$, according to the following equations:
\begin{eqnarray}
	\label{C_binned} C_{i} &=& \frac{n_{r,i}}{n_{d,i}}  \\
	\label{R_binned} R_{i} &=& \frac{n_{r,i}}{n_{t,i}}
\end{eqnarray}

Parameter correctness describes how accurate the source finder is in calculating the parameters of the sources it finds. These parameters include not only the position and size of the source, in the spatial and spectral dimensions, but also properties such as peak flux. The accuracy for the parameterisation may be described in terms of the difference between the real value of a source, and the corresponding value measured by the source finder.

These accuracy measures are calculated by executing the source finder on data cubes for which the catalogue of objects is already known, and comparing this catalogue to the catalogue produced by the source finder. For the purposes of this paper, the known source catalogue shall be called the \emph{reference catalogue} and the catalogue produced by the source finder shall be called the \emph{test catalogue}, respectively. The accuracy of a source finder is defined relative to the reference catalogue. That is, the reference catalogue is assumed to contain exactly the list of objects that are in the corresponding data cube. For example, the reference catalogue may be a list of sources used to create a simulation, a list of artificial sources added to a real data cube, or a list of sources found by a previous examination of the data cube. The following section will describe the method used by SFAE to cross-match a reference catalogue to a test catalogue.

\section{Method}
\begin{figure}
	\centering
	\includegraphics[width=0.5\columnwidth]{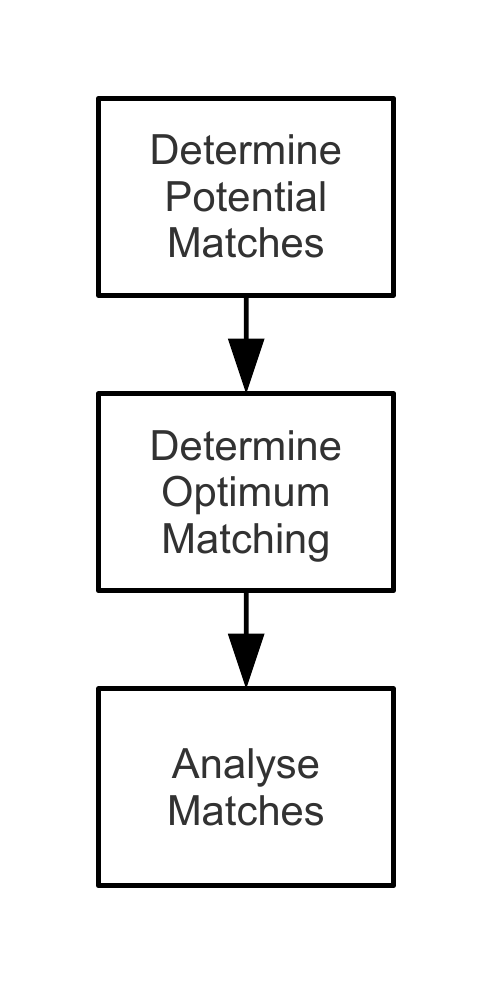}
	\caption{{Process flowchart.} This diagram shows the steps in determining the accuracy of a source finder for a particular data cube. The sources are first compared to determine potential matches, then the potential matches are sorted into final matches, and finally these final matches are analysed to determine the source finder's accuracy.}
	\label{process_flowchart}
\end{figure}

This section describes the method by which SFAE measures the accuracy of a source finder. SFAE requires several steps to analyse the accuracy of a source finder, as shown in \figurename\ \ref{process_flowchart}. The first step is to compare the position and frequency of each reference object to those of each test object, and the differences between these are used to place the objects into potential matches. Then, if there are any sets of potential matches that are not one-to-one, these sets of objects and their distances are used to determine an optimum matching between reference objects and test objects. Finally, the matched sources may be analysed to determine the accuracy of the test catalogue, as a function of the properties of the sources. Each of these steps will now be described in greater detail.

\subsection{Step One: Potential Matches}
\begin{figure}
	\centering
	\includegraphics[width=\columnwidth]{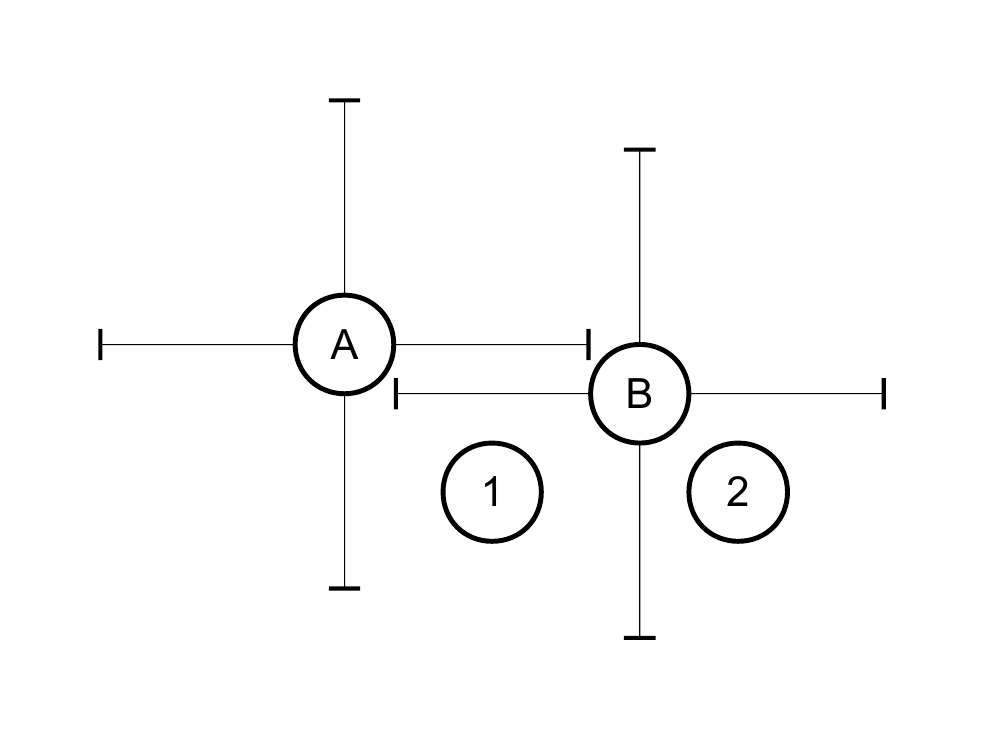}
	\caption{{Confused Sets.} It is possible for objects to be in more than one potential match. Suppose that objects A and B are reference sources, and objects 1 and 2 are test sources. The lines show the thresholds used to determine potential matches. These objects would be placed in the same a confused set, and a more sophisticated algorithm is needed to determine which reference object matches which test object.}
	\label{confused_sets}
\end{figure}

The first step of comparing reference objects to test objects is to determine which pairs could possibly refer to the same object in the data cube. SFAE applies a function that determines whether a given reference source and test source are close enough to be considered potentially the same object. Each pair of reference and test objects that has been accepted by the potential match function are deemed a \emph{potential match}. If a reference source and a test source have only each other as potential matches, they can be marked as a final match. If a source from either catalogue has no potential matches, it is considered an unmatched source. If one or both of them have more than one potential match, then a set of objects is created containing the pair of objects. The potential matches of the objects inside the set are then added to the set, until all the potential matches of all the objects in the set are themselves in that set. Such a set of objects is called a \emph{confused set}, an example of which is shown in \figurename\ \ref{confused_sets}. The objects in each confused set are sorted into their final matches in the next stage of the algorithm.

The function used by SFAE applies three thresholds to the properties of a reference-test object pair to determine if they are a potential match of each other. The first two thresholds check if the spatial distance between the two objects is less than or equal to the size of the telescope beam, along the beam's major and minor axes respectively. The third comparison is applied to the frequency of the entries, where the two sources are considered close enough if the difference between the frequency of the two entries is less than or equal to the full width at half maximum (FWHM) of the catalogue source. Ideally this function should be based on the error of the test detections relative to the reference catalogue being considered, but the size of the error may not be known. If such information is available about the error between the two catalogues, an alternative potential match function that incorporates this information may be used instead.

\subsection{Step Two: Optimum Matching}
The second step of cross-matching the reference and test catalogues is to make final pairs from the confused sets formed in the previous step. This step enforces a matching scheme where each object is either in a single match, or no matches at all. Additionally, for any reference object $a$ that is matched to test object $b$, $b$ is also matched to object $a$. Final matches are only allowed between reference-test object pairs that have been previously marked as potential matches, to ensure that SFAE only makes valid matches.

Two particular subsets of confused sets that are of interest are the cases of merged or split objects. Respectively, this is where the source finder has taken two sources and reported them as a single object and where the source finder has found a single source and reported it as multiple objects, resulting in a reduced completeness. In the case of a merged object, this may appear to SFAE as a confused set with two or more reference objects and a single test object. It is difficult for SFAE to determine whether such a test object genuinely matches all the corresponding reference sources, or if there are reference sources that have not been found. SFAE uses the reference catalogue as the truth, so it will respond to this situation by matching one of the reference objects to the test object, and leaving the other reference objects unmatched. Leaving the extra sources unmatched conveys the information that the source finder has incorrectly merged the two objects. It is possible to account and correct for this by examining the merged test source and its potential matches, and using their parameters to manually decide whether or not the reference sources are all part of the test source. The same applies, in reverse, to split objects.

The optimum matching is found by employing a more fine-grained technique, the \emph{Hungarian Algorithm} \citep{HungarianAlgorithm}. The Hungarian Algorithm considers the entries as a weighted bipartite graph, where each vertex in the graph corresponds to either a reference object or a test object. The matching produced by this algorithm is one that has the maximum number of matches, and among all the possible ways this matching can be achieved, it picks the one with the least total distance between the paired objects. Edges exist between any and all pairs of nodes that have been identified as potential matches in step two.

The weights between the nodes are dimensionless values that define how different the reference and test objects are to each other. If $\Delta a$ and $\Delta b$ are the distances between the two objects along the major and minor axes of the beam respectively, $w_{a}$ and $w_{b}$ are the widths of the beam along the major and minor axes of the beam, $\Delta f$ is the difference in frequency position between the two objects, and $w_{f}$ is the frequency resolution of the telescope or simulation used to create the data cube, then the distance weighting function between two objects $x$ and $y$ is defined as $\delta_{x, y}$:
\begin{equation}
	\label{DistanceMetric}
	\delta_{x, y} = \sqrt{\left(\frac{\Delta a}{w_{a}}\right)^{2} + \left(\frac{\Delta b}{w_{b}}\right)^{2} + \left(\frac{\Delta f}{w_{f}}\right)^{2}}
\end{equation}
If the frequency resolution is unknown, then it may be possible to approximate $w_{f}$ with the value of the channel width.

Because only potential matches are being compared by the distance metric function, the frequency difference between different sources is small and therefore the difference in frequencies is a good approximation for the difference in distance between the objects. It would also be possible to include other parameters in the distance function, such as peak flux or frequency width, which would cause the algorithm to favour matching objects with similar parameters.

However, the decision was made to not use other parameters in this equation because doing so would cause errors in the parameterisation of the test objects to affect the matching of the reference and test objects. Additionally, incorporating other parameters would require an appropriate weighting scheme to ensure that all the parameters are fairly and accurately considered in the distance between the two objects. The weighting function would need to account for the systematic and random errors for each parameter used to calculate the distance between two objects. The distance function is limited to just the position and frequency because together these three parameters can uniquely identify a source and because they are the core function of a source finder: to identify the objects in a data cube and report their positions.

\subsection{Step Three: Analyse Matches}
Now that it is known what reference sources correspond to what test sources, it is possible to analyse these matches in order to calculate the accuracy of the source finder being tested. The completeness values can be calculated using Equations \ref{C_binned} and \ref{R_binned}, setting $n_{r,i}$ to the number of matched sources that are in the $i$th bin. For consistency, the parameters used to determine what bin a particular source is in should come from the reference catalogue when calculating completeness and the test catalogue when calculating the reliability. The value of $n_{d,i}$ is the number of objects in the reference catalogue, both matched and unmatched, that are in the $i$th bin and $n_{t,i}$ is the number of test objects, both matched and unmatched, that are in bin $i$. The accuracy of the parameterisation of the source finder can be measured by taking each pair of matched reference-test sources, and comparing the parameter of the test source to the value of the reference source.

In addition to calculating the completeness and reliability of a source finder, these matches may also be used to determine the difference between the sources that were and were not found by the source finder, and the true and false detections of the source finder. The differences for one or more parameters can be shown by plotting the distribution of the matched and unmatched test sources, as a function of these parameters. This information can be used to identify what types of sources are missed by the source finder. The differences between true and false detections can be used to suggest selection functions that could be applied to the catalogue produced by the source finder, removing suspected false detections in order to improve its reliability.

This section has shown how SFAE program determines the accuracy of a source finder. If there are $N$ objects in the reference catalogue list for the source cube, then reading in the data has a time complexity of $O(N)$. Comparing the reference objects to the test objects in Step one has a time complexity of $O(N^{2})$. The optimum matching algorithm in Step two has a time complexity of $O(N^{3})$ \citep{NetworkProblemImprovements}. The time complexity of Step three varies with the type of data analysis done. Calculating the completeness and reliability as a function of source parameter has a time complexity of $O(N \log N)$ and the parameter comparison analyses each have a time complexity of $O(N)$. Therefore, this program has an overall time complexity of $O(N^{3})$. The information produced by SFAE is demonstrated in the following section.

\section{Results}
The SFAE program was tested, both to ensure that it correctly measures the accuracy of a source finder and to demonstrate the information that it can give. The test data set is based on a FITS format ASKAP simulation data cube \citep{MWhitingReducedNoise}. This simulated data cube covers an area $1.42 \times 1.42$ degrees with an angular resolution of $30$ arcsec, a pixel size of $10$ arcsec and a frequency range of $1327.39$ to $1422.0175$ MHz. The frequency resolution and channel width are both $92.5$ kHz. This results in a data cube composed of $512 \times 512 \times 1,024$ voxels. This data cube is intended for use in testing source finders, so it has artificially reduced noise, in order to increase the number of sources visible to the source finder. The source catalogue used as the input for the simulation is from the SKADS S3-SAX simulation \citep{SKADSSimulation}. The sources used are mostly point sources but there are some extended sources. The test data cube is supplied as a dirty image, so prior to running the source finder the image was cleaned using \textsc{miriad} \citep{MiriadUserGuide}, using the full PSF data cube.

This cleaned image still has a number of sidelobes, from two sources. The first source of sidelobes is from errors in the cleaning process, as the PSF cube provided does not exactly match the PSF sidelobes present in the dirty image, due to what is suspected to be an error in the pipeline creating this simulated data. The second is that there are a number of sidelobes from sources that are outside the data cube, and therefore cannot be removed by the \textsc{clean} task. The sidelobes cause a number of erroneous detections, as the source finder tested finds peaks in the sidelobe patterns. In the general case, data sets may contain errors from their construction, such as telescope noise, pipeline errors, simulation errors, and so on. Source finders will need to be able to operate on data sets with such errors. When using SFAE to determine the accuracy of a source finder, the data cube should be given to the source finder in the same manner as the source finder would receive data cubes in a production environment.

The source finder being tested is \textsc{Duchamp} \citep{Duchamp_prepaper}, version $1.1.13$. However, the information given here should be considered a demonstration of SFAE rather than a test of the accuracy of \textsc{Duchamp}. The test catalogue was obtained by executing \textsc{Duchamp} on the cleaned data cube. \textsc{Duchamp} uses a number of parameters to calculate its list of detections, in addition to the data cube. The parameters specified to create the detection list used for this test use the default parameters, with the exception of using \`a trous wavelet reconstruction in three dimensions. \textsc{Duchamp}'s default parameter set uses a three sigma threshold for the final objects.

The reference catalogue list was obtained by running \textsc{Duchamp} on a corresponding model image of the dirty cube. This model data cube contains only the sources present in the test data cube, with zero noise. In obtaining the reference catalogue, the default parameters for \textsc{Duchamp} were used except for using a direct flux threshold of $1\:\upmu$Jy. It can be reasonably assumed that the source finder derives the correct parameters from the model cube, and it is how the source finder parameterises the noisy image that is of interest. Whether or not a source finder does, in fact, correctly parameterise a model image can be determined by using several different source finders to search and parameterise the model image, then comparing their results to see if they calculate the same values.

The reference catalogue has $235$ entries and the test catalogue has $384$ entries. Of these, there are $190$ matched pairs, using these values with equations \ref{CEq} and \ref{REq} results in an overall completeness of $70.6\%$ and a reliability of $49.9\%$. The one-to-one matching did not exclude any sources from being matched.

\begin{figure}
	\centering
	\resizebox{\columnwidth}{!}{\input{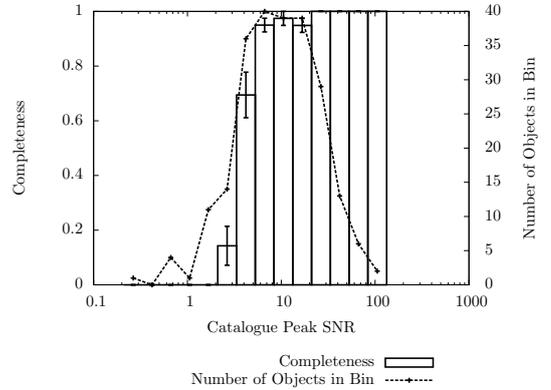}}
	\caption{{Completeness by Peak Flux.} This graph shows the completeness of the test source finder as a function of the peak SNR of an object. The abscissa is the peak SNR range of that bin, with a logarithmic scale. The histogram shows the completeness of the source finder, for that bin. The error bars show a one standard deviation error, as determined by bootstrap resampling. The dotted line shows the number of objects in each bin. This graph uses the reference object's value for the peak flux of a source.}
	\label{completeness_by_peak_flux}
\end{figure}

\begin{figure}
	\centering
	\resizebox{\columnwidth}{!}{\input{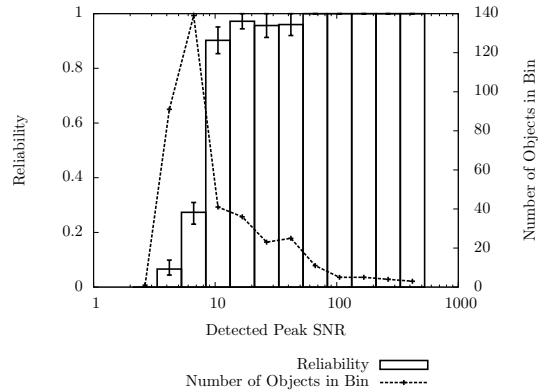}}
	\caption{{Reliability by Peak Flux.} This graph shows the reliability of the test source finder as a function of the peak flux of an object. The abscissa is the peak flux range of that bin, with a logarithmic scale. The histogram shows the reliability of the source finder, for that bin. The error bars show a one standard deviation error, as determined by bootstrap resampling. The dotted line shows the number of objects in each bin. This graph uses the test object's value for the peak flux.}
	\label{reliability_by_peak_flux}
\end{figure}

The completeness of the source finder as a function of the peak SNR of a source is shown in \figurename\ \ref{completeness_by_peak_flux}, using Equation \ref{C_binned}. The SNR of a source is calculated by dividing the peak flux of a source, in units of Jy per beam, by the rms noise, as calculated from the cleaned data cube using the \textsc{miriad} routine \textsc{histo}. Likewise, the reliability of the test source finder as a function of peak SNR is shown in \figurename\ \ref{reliability_by_peak_flux}, using Equation \ref{R_binned}.

The error bars were determined using \emph{bootstrap resampling}. For each bin, an object is randomly selected from the set of objects in that bin and it is recorded whether or not it has a matching test source. This procedure is repeated a $N$ times, with $N$ being the number of objects in the  bin considered. Using Equation \ref{C_binned}, the completeness is then calculated from the objects that had been selected. A total of $1\,000$ completeness values are calculated per bin, and the $15.9\%$ and $84.1\%$ percentile completeness values were used as the lower and upper standard deviation, respectively. Note that if the sources in a particular bin are either all matched or all unmatched, then that bin will always have the same completeness (either 100\% or 0\%), no matter which objects are randomly selected, and therefore that bin will not have any error bars plotted.

\begin{figure}
	\centering
	\subfigure[{Reference Sources by Integrated Flux and Line Width}]{
	\resizebox{\columnwidth}{!}{\input{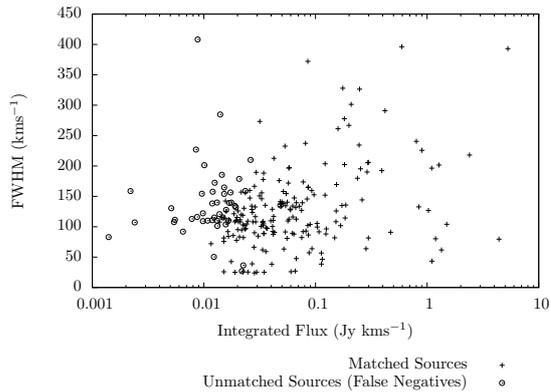}}
	%\caption{ }
	\label{cat_int_flux_vel_scatter_plot}
	}
	\subfigure[{Test Sources by Integrated Flux and Line Width}]{
	\resizebox{\columnwidth}{!}{\input{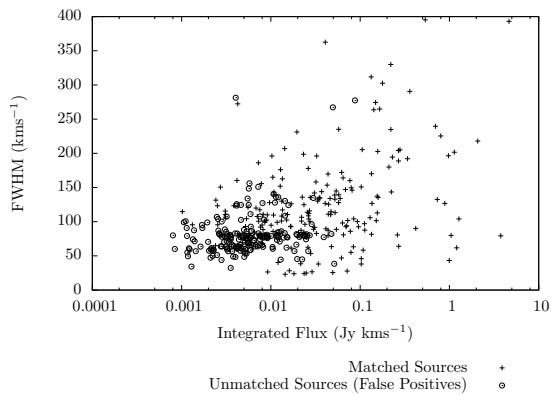}}
	\label{det_int_flux_vel_scatter_plot}
	}
	\caption{ These graphs show the distribution of matched and unmatched sources as a function of their integrated flux and line width. The abscissa is the integrated flux of the source and the ordinate is the FWHM of the source. The first plot shows the completeness of the source finder using the reference sources. The second plot shows the reliability of the source finder, using the test parameters.}
	\label{int_flux_vel_scatter_plot}
\end{figure}

The distribution of the reference and test sources as a function of their integrated flux and line width is shown in \figurename\ \ref{int_flux_vel_scatter_plot}. The matched sources are those that are both present in the reference catalogue and have been found by the source finder. The unmatched reference sources are those that have not been found by the source finder, and the unmatched test sources are false detections. The abscissa is the integrated flux of a source, and the ordinate is the FWHM size of the sources, using the reference catalogue parameters.

\begin{figure}
	\centering
	\resizebox{\columnwidth}{!}{\input{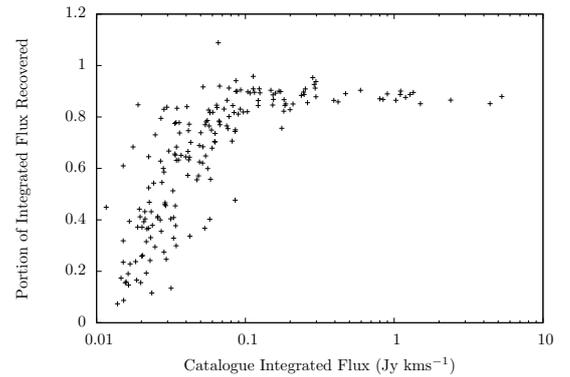}}
	\caption{{Integrated Flux Parameter Comparison.} This graph shows how accurately the source finder determines the integrated flux of sources. The abscissa is the reference catalogue integrated flux of a matched source. The ordinate is the portion of the integrated flux of the source found by the source finder.}
	\label{integrated_flux_comparison}
\end{figure}

The matches produced by SFAE can be analysed to determine how accurately the source finder parameterises the sources it finds. One such analysis can be seen in \figurename\ \ref{integrated_flux_comparison}. This compares how well the test source finder measures the integrated flux of a source. The abscissa is the integrated flux of a source in Jy kms\textsuperscript{-1}, as listed in the reference catalogue. The ordinate is the portion of the integrated flux of the source that was detected by the source finder.

The accuracy of SFAE itself was measured by manually matching entries from the reference catalogue to the entries in the test catalogue list. This was done by plotting the positions of the reference and test objects. Then each reference object's position was observed and it was recorded either which test object was closest to the reference object in question, or that the reference object had no nearby test object. The limits used were the size of the major axis of the beam in spatial distance, and the resolution of the data cube, in frequency. Upon comparing this manual matching to the one produced by the SFAE algorithm, SFAE matched each of the $235$ objects in the reference catalogue to the same test object as the manual matching.

\section{Discussion}
The results of the SFAE program show a variety of information about the performance of the example source finder \textsc{Duchamp} and the parameter set used. The overall completeness and reliability figures show how successful the test source finder was overall in correctly locating the sources in the cube. More useful are the completeness and reliability values plotted as a function of different parameters. \figurename\ \ref{completeness_by_peak_flux} shows the portion of sources found as a function of the peak flux. This plot can show for which peak flux value the source finder can be considered complete, before the rate at which the source finder detects objects drops off. The reliability of the source finder as function of peak flux is shown in \figurename\ \ref{reliability_by_peak_flux}. This analysis shows at what peak flux value the source finder can be considered to find primarily genuine detections, rather than false ones. Together, the information in these graphs can be used, for example, to determine the flux limit of the survey the source finder is used for.

SFAE's ability to analyse a source finder's completeness and reliability is further demonstrated in \figurename\ \ref{int_flux_vel_scatter_plot}. These plot shows how reference and test sources are distributed as a function of both their integrated flux and their FWHM. \figurename\ \ref{cat_int_flux_vel_scatter_plot} shows the difference between the sources that have and have not been found by the source finder. This information can be used to identify potential areas of improvement to the searching algorithm and for what types of sources the produced catalogue will be incomplete. The differences between true and false detections can be seen in \figurename\  \ref{det_int_flux_vel_scatter_plot}. To the extent that the test data cube represents real observation data, these differences can be used to determine a selection function to apply to the source finder's output, in order to improve its reliability.

The objects and their matches can be used to compare how well the source finder determines the parameters of the sources it finds. \figurename\ \ref{integrated_flux_comparison} shows what the source finder calculated as each object's integrated flux, compared to what was listed in that object's reference catalogue entry. This figure shows that the source finder calculated the integrated flux of almost all of its sources as lower than the actual integrated flux, and that the reduction in the listed flux increases as the reference value for the integrated flux decreases. The error shown suggests that the source finder should improve its existing parameterisation, use a new algorithm, or use this error information to devise a statistical correction to apply. The errors in the parameterisation should also be kept in mind when assessing the above measures of accuracy as a function of the parameters reported by the source finder. Similar analyses can be run on other parameters to determine the characteristic of the error in the source finder's analysis of that parameter.

The overall time complexity for SFAE is $O(N^{3})$, where $N$ is the number of objects in the data cube catalogue. Running SFAE for the $235$ objects in the ASKAP simulated data cube resulted in a running time of less than three seconds, on a single quad-core CPU system. As the running time for \textsc{Duchamp} on the same system for the test data cube is $90$ minutes, over three orders of magnitude longer, the running time of SFAE can be considered negligible in comparison. On the scale of ASKAP, the Widefield ASKAP Legacy L-band Blind All-sky surveY (WALLABY) is expected to find approximately $500\: 000$ galaxies across an estimated area of $3\pi$\:sr \citep{WallabyProposal}. With ASKAP's instantaneous field of view of approximately $30\:\deg^{2}$ there will be around $1\: 200$ separate fields for the entire survey \citep{WALLABYSkyTileEstimate} and therefore approximately $500$ sources per data cube. From the recorded running time of $2.138$\:s for $2\, 345$ objects and the time complexity above this suggests a running time on the order of $18$ s for a single ASKAP-scale data cube. Therefore, no further attempts to improve the execution speed are necessary for the scale of data currently being used.

\section{Summary}
The Source Finder Accuracy Evaluator provides an automated, deterministic method to determine the accuracy of a source finder. The program does this by taking the results of the source finder in question from a data cube with a known source list, and comparing the results against a known source catalogue. Using the two catalogues of sources, and header information from the data cubes, SFAE creates a list of which reference object-test object pairs refer to the same source, if any.

The list of matches produced by SFAE may then be analysed to determine the completeness and reliability of the source finder. The completeness and reliability figures can be calculated for both the catalogues as a whole, and as a function of the parameters of the sources in the catalogues. Breaking this information down by the properties of the sources allows SFAE to provide a more detailed account of the accuracy of the source finder, characterising the types of sources the source finder can accurately find, and the types of sources it either misses or erroneously locates.

Finally, the accuracy of the source finder's para\-meterisation algorithms is determined by comparing the value of the selected parameter from each source, as reported by the source finder, against the same values as recorded in the reference catalogue list. Therefore, SFAE provides a variety of information useful to determining the accuracy of source finders, both in terms of the sources they find, or don't find, in addition to how accurately they characterise these sources.

\subsection{Future Work}
	SFAE could be modified to run its analysis over multiple data cubes and their corresponding source lists, to collate source finder accuracy data from a large test data set. SFAE could also use a more sophisticated approach to dealing with instances where the source finder splits a reference catalogue source into two or more separate test objects, or when the source finder merges multiple objects into a single test source. In the future, SFAE and its results could be used to provide feedback for computer learning-based source finding algorithms, and used to train them.
	
	As a special case, it may be possible to extend this work for use in cross-matching catalogues, with certain alterations. This application is beyond the scope of this work, and the method may not provide complete matching, as matching between sources in different frequencies may not have a one-to-one relation to each other. For example, a single detection in one band may match multiple object in another band. The changes necessary would involve using different threshold and distance functions. These would need to use distance, red shift or velocity in place of frequency. The thresholds applied to determine potential matches and the weightings applied to different parameters in the distance function would need to be set depending on the relative errors particular to the catalogues being matched.

\section*{Acknowledgement}
The authors thank Matthew Whiting for writing the \textsc{Duchamp} source finder and creating the ASKAP simulation data used in this work. The authors also thank Attila Popping for his feedback in preparing this work.

%Bibliography macro(s)
\let\jnl@style=\rm
\def\ref@jnl#1{{\jnl@style#1}}
\def\mnras{\ref@jnl{MNRAS}}

% that's all folks

\begin{thebibliography}{}
\expandafter\ifx\csname natexlab\endcsname\relax\def\natexlab#1{#1}\fi

\bibitem[{{Barnes} {et~al.}(2001){Barnes}, {Staveley-Smith}, {de Blok},
  {Oosterloo}, {Stewart}, {Wright}, {Banks}, {Bhathal}, {Boyce}, {Calabretta},
  {Disney}, {Drinkwater}, {Ekers}, {Freeman}, {Gibson}, {Green}, {Haynes}, {te
  Lintel Hekkert}, {Henning}, {Jerjen}, {Juraszek}, {Kesteven}, {Kilborn},
  {Knezek}, {Koribalski}, {Kraan-Korteweg}, {Malin}, {Marquarding}, {Minchin},
  {Mould}, {Price}, {Putman}, {Ryder}, {Sadler}, {Schr{\"o}der}, {Stootman},
  {Webster}, {Wilson}, \& {Ye}}]{HIPASSDataReduction}
{Barnes}, D.~G., {et~al.} 2001, \mnras, 322, 486

\bibitem[{DeBoer {et~al.}(2009)DeBoer, Gough, Bunton, Cornwell, Beresford,
  Johnston, Feain, Schinckel, Jackson, Kesteven, Chippendale, Hampson,
  O'Sullivan, Hay, Jacka, Sweetnam, Storey, Ball, \& Boyle}]{ASKAPDataOutput}
DeBoer, D., {et~al.} 2009, Proceedings of the IEEE, 97, 1507

\bibitem[{Edmonds \& Karp(1972)}]{NetworkProblemImprovements}
Edmonds, J., \& Karp, R.~M. 1972, J. ACM, 19, 248

\bibitem[{Koribalski \& Staveley-Smith(2009)}]{WallabyProposal}
Koribalski, B.~S., \& Staveley-Smith, L. 2009, Proposal for WALLABY: Widefield
  ASKAP L-band Legacy All-sky Blind surveY, Available from
  \url{http://www.atnf.csiro.au/research/WALLABY/proposal.html}, Last Visited 9
  September 2011

\bibitem[{Kuhn(1955)}]{HungarianAlgorithm}
Kuhn, H. 1955, Naval research logistics quarterly, 2, 83

\bibitem[{Meyer {et~al.}(2004)Meyer, Zwaan, Webster, Staveley-Smith,
  Ryan-Weber, Drinkwater, Barnes, Howlett, Kilborn, Stevens,
  {et~al.}}]{HIPASSCatalogue}
Meyer, M., {et~al.} 2004, \mnras, 350, 1195

\bibitem[{Obreschkow {et~al.}(2009)Obreschkow, Kl{\"o}ckner, Heywood, Levrier,
  \& Rawlings}]{SKADSSimulation}
Obreschkow, D., Kl{\"o}ckner, H.-R., Heywood, I., Levrier, F., \& Rawlings, S.
  2009, The Astrophysical Journal, 703, 1890

\bibitem[{Sault \& Killeen(2008)}]{MiriadUserGuide}
Sault, B., \& Killeen, N. 2008, Miriad Multichannel Image Reconstruction, Image
  Analysis and Display Users Guide, Available from
  \url{http://www.atnf.csiro.au/computing/software/miriad/}, Last Visited 15
  December 2010

\bibitem[{Warren(2011)}]{WALLABYSkyTileEstimate}
Warren, B.~E. 2011, Sky Tessellation Patterns for Field Placement for the
  All-Sky HI Survey WALLABY, Tech. rep.

\bibitem[{Westerlund(2010)}]{DuchampParallel}
Westerlund, S. 2010, iVEC Internship Report, Available from
  \url{http://www.icrar.org/__data/assets/pdf_file/0006/1750866/stefan_westerl%
und_ivec_report.pdf}, Last Visited 21 July 2011

\bibitem[{Whiting(2008)}]{Duchamp_prepaper}
Whiting, M. 2008, Galaxies in the Local Volume, ed. B.~Koribalski \& H.~Jerjen,
  Astrophysics and Space Science Reviews (Springer), 343--344

\bibitem[{Whiting(2010)}]{MWhitingReducedNoise}
Whiting, M. 2010, available from:
  \url{http://www.atnf.csiro.au/people/Matthew.Whiting/ASKAPsimulations}, Last
  Visited 15 December 2010

\bibitem[{{Zwaan} {et~al.}(2004){Zwaan}, {Meyer}, {Webster}, {Staveley-Smith},
  {Drinkwater}, {Barnes}, {Bhathal}, {de Blok}, {Disney}, {Ekers}, {Freeman},
  {Garcia}, {Gibson}, {Harnett}, {Henning}, {Howlett}, {Jerjen}, {Kesteven},
  {Kilborn}, {Knezek}, {Koribalski}, {Mader}, {Marquarding}, {Minchin},
  {O'Brien}, {Oosterloo}, {Pierce}, {Price}, {Putman}, {Ryan-Weber}, {Ryder},
  {Sadler}, {Stevens}, {Stewart}, {Stootman}, {Waugh}, \&
  {Wright}}]{HIPASSAccuracy}
{Zwaan}, M.~A., {et~al.} 2004, \mnras, 350, 1210

\end{thebibliography}
\end{document}